\documentclass[pre,aps,superscriptaddress,showpacs]{revtex4}

\def \beq{\begin{equation}}         \def \eeq{\end{equation}}
\def \beqa{\begin{eqnarray}}        \def \eeqa{\end{eqnarray}}
\def \bea{\begin{array}}        \def \eea{\end{array}}

\usepackage{amsmath}
\usepackage{epsfig}
\usepackage[dvips]{color}

\begin{document}

\title{Operator perturbation theory in the backward Heisenberg picture }
\author{Fei Liu }
 \email[Email address: ]{feiliu@buaa.edu.cn}
\author{Ding-yang Liu}
\affiliation{School of Physics and
Nuclear Energy Engineering, Beihang University, Beijing 100191,
China}
\date{\today}

\begin{abstract}
{We present a simple operator perturbation theory in the
backward Heisenberg picture. Compared with the well-known
Heisenberg picture, the revised picture is based on the
backward time instead of the forward time. The unique feature
of the uncommon picture is that the perturbation expansion
becomes very simple and the famous Dyson expansion is not
directly involved. Its relationship with the perturbation
expansion for the density operator developed by Kubo is also
discussed.}
\end{abstract}
 \maketitle

{\it Introduction.} Consider a perturbed quantum system with
the Hamiltonian
\begin{eqnarray}
\label{perturbedHamiltonian}
H(t)=H_0+ H_1(t),
\end{eqnarray}
where $H_0$ is the time-independent part and $H_1$ is a small
time-dependent perturbation which is assumed to be switched on
at initial time 0. We are interested in the following question:
given an observable $F$ that does not explicitly depend on time
and its Heisenberg picture defined as
\begin{eqnarray}
F^{H}(t)=U^{\dag}(t)F U(t),\label{Hspicture}
\end{eqnarray}
where $U(t)$ is the time evolution operator of the Hamiltonian
$H(t)$ and the superscript $H$ stands for the Heisenberg
picture, can we expand the dynamic observable in the order of
the perturbation based on the unperturbed system without
directly resorting to the famous Dyson expansion~\cite{Dyson}?
Apparently, if the constraint was dropped, the problem would
become almost trivial because of
\begin{eqnarray}
U(t)&=&U_0(t){\cal T}_{+}e^{{(i\hbar)}^{-1}\int_{t'}^t
U^\dag_0(\tau)H_1(\tau)U_0(\tau)d\tau }
\end{eqnarray}
where ${\cal T}_{+}$ is the time-ordering operator, and
$U_0(\tau)$$=\exp(H_0t/i\hbar)$ is the time-evolution operator
for the unperturbed system. This question was raised when we
investigated the fluctuation relations in nonequilibrium
processes~\cite{Bochkov77,Evans93,JarzynskiPRL97,TalknerPRE07,QuantumLiuF},
and in the work we give a definite answer. We must emphasize
that we aim to reinvestigate the old operator perturbation
issue from a very different viewpoint and are not intended to
present a method as a substitution for the Dyson expansion. In
fact, we cannot even completely exclude that the
same method has been in the literature. \\

{\it The backward Heisenberg picture.} We start by introducing
a revised Heisenberg picture observable~\cite{QuantumLiuF}
\begin{eqnarray}
F^{B}(t,t')=U(t')F^H(t)U^{\dag}(t')\label{Kspicture}
\end{eqnarray}
with $0\leq t'\leq t$. Because $t'$ is the {\it backward} time,
we call the picture as the backward Heisenberg picture and
specifically indicate it by a superscript $B$. Obviously, if
one selects $t'$=0 and $t$, the picture reduces to the
Heisenberg and Schr\"{o}dinger pictures, respectively. The
equation of motion for the new defined dynamic observable
$F^{B}(t,t')$ with respect to $t'$ is simply
\begin{eqnarray}
\label{Kseqmotion}
i\hbar\partial_{t'}F^B(t,t')=[H(t'),F^B(t,t')].
\end{eqnarray}
It is worthy emphasizing that this is a terminal condition
problem with $F^B(t,$ $t)$$=$$F$ and that the Hamiltonian
operator therein is indeed the Schr\"{o}dinger picture, which
is significantly different from the standard Heisenberg
equation of motion,
\begin{eqnarray}
i\hbar \partial_t F^H(t)=-[H^H(t),F^{H}(t)].\label{Hseqmotion}
\end{eqnarray}
with the initial condition $F^{H}(0)=F$. The careful reader may
notice that Eq.~(\ref{Kseqmotion}) is almost same with the
equation of motion for the density matrix $\rho(t)$ except that
$t'$ here is in place of the forward time $t$, which is not
accidental because of
\begin{eqnarray}
\partial_{t'} {\rm Tr}\{F^B(t,t')\rho(t')\}=0.
\end{eqnarray}
The expectation value $\langle F\rangle(t)$ of the operator at
time $t$ in the picture is
\begin{eqnarray}
{\rm Tr}\{F\rho(t)\}={\rm Tr}\{F^{H}(t)\rho_0\}={\rm Tr}\{F^{B}(t,0)\rho_0\}, \label{expectationvalule}
\end{eqnarray}
where $\rho_0$ is the initial density operator that may be a
pure or mixed state. To illustrate the uncommon picture, we
calculate the position and momentum's backward Heisenberg
pictures for a one-dimensional harmonic oscillator with a
Hamiltonian $H=p^2/2m+m\omega^2 x^2/2$. Because the system is
time-independent and one may easily find their equations given
by
\begin{eqnarray}
\partial_{t'}x(t,t')&=&-p(t,t')/m, \\
\partial_{t'}p(t,t')&=&m\omega^2 x(t,t').
\end{eqnarray}
For simplicity in notations, we did not write the superscript
$B$ out explicitly. Solving both equations with the terminal
conditions $x(t,t)=x$ and $p(t,t)=p$, we obtain
\begin{eqnarray}
x(t,t')&=&[x\sin \omega t- \frac{p}{m\omega}\cos\omega t]\sin\omega t'+[x\cos\omega t+\frac{p}{m\omega}\cos\omega t]\cos\omega t',\\
p(t,t')&=&-[m\omega x\sin \omega t- p \cos\omega t]\cos \omega t'+[m\omega x\cos\omega t+p\cos\omega t]\sin\omega t'.
\end{eqnarray}
Letting $t'=0$, the above solutions reduce to those directly
calculated in the Heisenberg picture~\cite{Sakurai}. In fact,
if the Hamiltonian does not depend on time, the backward
Heisenberg picture~(\ref{Kspicture}) could not possibly show
any distinctive features compared with the standard Heisenberg
picture because of obvious
\begin{eqnarray}
F^{B}(t,t')=F^{H}(t-t').
\end{eqnarray}

{\it Operator's perturbation}. Now we turn our attention to the
perturbation case~(\ref{perturbedHamiltonian}). Substituting
the perturbation expansion
\begin{eqnarray}
F^{B}(t,t')=F^{B}_0(t,t')+F^{B}_1(t,t')+F^{B}_2(t,t')+\cdots
\end{eqnarray}
into the equation of motion~(\ref{Kseqmotion}) and comparing
its both sides order by order, we find
\begin{eqnarray}
\label{expanding0order}
&&\left\{%
\begin{array}{ll}
\i\hbar\partial_{t'} F^{B}_0(t,t')=[H_0,F^B_0(t,t')] \\
F^{B}_0(t,t)=F,\\
   \end{array}
\right.\\
\label{expanding1order}
\nonumber\\&&\left\{%
\begin{array}{ll}
\i\hbar\partial_{t'} F^{B}_1(t,t')=[H_0,F^{B}_1(t,t')]+[H_1(t'),F^{B}_0(t,t'),]\\
F^{B}_1(t,t)=0, \\
   \end{array}
\right.\\
\nonumber\\
\label{expanding2order}
&&\left\{%
\begin{array}{ll}
\i\hbar\partial_{t'} F^{B}_2(t,t')=[H_0,F^{B}_2(t,t')]+[H_1(t'),F^{B}_1(t,t')]\\
F^{B}_2(t,t)=0,
   \end{array}
\right.\\
&&\hspace{2cm} \cdots.\nonumber
\end{eqnarray}
Their formal solutions are
\begin{eqnarray}
F^{B}_0(t,t')&=&U_0(t')F^H_0(t)U^{\dag}_0(t')=F^H_0(t-t'),\label{solution0order}\\
F^{B}_1(t,t')&=&-(i\hbar)^{-1}U_0(t')\int_{t'}^t U^{\dag}_0(\tau')[H_1(\tau'),F^B_0(t,\tau')]U_0(\tau')d\tau'
U^{\dag}_0(\tau'),\label{solution1order}\\
F^{B}_2(t,t')&=&-(i\hbar)^{-1}U_0(t')\int_{t'}^t  U^{\dag}_0(\tau')[H_1(\tau'),F^B_1(t,\tau')]U_0(\tau')d\tau'
U^{\dag}_0(t'),\label{solution2order}\\
&&\cdots\nonumber
\end{eqnarray}
respectively. The simplicity of
Eqs.~(\ref{expanding0order})-(\ref{expanding2order}) contrasts
sharply with the complexity of the perturbation equations for
the Heisenberg picture observable
\begin{eqnarray} F^{
H}(t)=F^{H}_0(t)+F^{H}_1(t)+F^{H}_2(t)+\cdots,
\end{eqnarray}
each term on the right hand side of which satisfies
\begin{eqnarray}
\label{Hexpanding0order}
&&\left\{%
\begin{array}{ll}
\i\hbar\partial_{t} F^{  H}_0(t)=-[H_0,F^{  H}_0(t)] \\
F^{  H}_0(0)=F,\\
   \end{array}
\right.\\
\label{Hexpanding1order}
\nonumber\\&&\left\{%
\begin{array}{ll}
\i\hbar\partial_{t} F^{  H}_1(t)=-[H_0,F^{ H}_1(t)]-[(H_1)^{ H}_0(t)+(i\hbar)^{-1}\int_0^t[H_0,(H_1)^{ H}_0(\tau)]d\tau,
F^{ H}_0(t)]\\
F^{ H}_1(0)=0, \\
   \end{array}
\right.\\
\label{Hexpanding2order}\nonumber\\
&&\left\{%
\begin{array}{ll}
\i\hbar\partial_{t} F^{ H}_2(t)=-[H_0,F^{  H}_2(t)]-[(H_1)^{  H}_0(t)+(i\hbar)^{-1}\int_0^t[H_0,(H_1)^{ H}_0(\tau)]d\tau,
F^{ H}_1(t)]\\
\hspace{2.cm}-[(i\hbar)^{-2}[\int_0^t\int_0^{\tau'}
(H_1)^{  H}_0(\tau')(H_1)^{  H}_0(\tau'')d\tau' d\tau'',H_0]_+
-(i\hbar)^{-1}[\int_0^t (H_1)^{ H}_0(\tau)d\tau,(H_1)^{  H}_0(t)]
\\
\hspace{2cm}-(i\hbar)^{-2}\int_0^t(H_1)^{  H}_0(\tau')d\tau'H_0\int_0^t(H_1)^{  H}_0(\tau'') d\tau''
,\hspace{0.1cm}F_0^{ H}(t)]\\
F^{  H}_2(0)=0,
   \end{array}
\right.\\
&&\hspace{2cm}\cdots,\nonumber
\end{eqnarray}
respectively, where $[$ $,$ $]_+$ denotes an anticommutator. We
expand them until the second order, because the evolution of
the $n$-th order $F_n^{H}(t)$ depends on all orders from 0 to
$n$, and the calculations of these coefficients become
considerably tedious as the order increases. However, we must
point out that the relation~(\ref{expectationvalule}) has
imposed conditions
\begin{eqnarray}
F^{H}_n(t)=F^{B}_n(t,0) \hspace{0.5cm} (n=0,1,2,\cdots).
\end{eqnarray}

In the Schr\"{o}dinger picture, Kubo has developed a method to
evaluate the perturbation expansion for the density matrix
\begin{eqnarray}
\rho(t)=\rho_0+\rho_1(t)+\rho_2(t)+\cdots
\end{eqnarray}
on the basis of its equation of motion in 1957, which does not
depend on the Dyson expansion as well~\cite{Kubo}. One may see
that our calculation above is highly analogous to his method.
Particularly, Eq.~(\ref{expectationvalule}) has indicated
\begin{eqnarray} {\rm Tr}\{\rho_n(t)F\}={\rm
Tr}\{F_n^B(t,0)\rho_0\}.\label{norderDensityMatrix}
\end{eqnarray}
It is interesting to concretely calculate how these identities
arise using the
solutions~(\ref{solution0order})-(\ref{solution2order}). We use
the first three order terms ($n=0,1,2$) as examples. The
identity with zero order is obviously trivial. For the first
and second order terms, we have
\begin{eqnarray}
\label{1stexpctation}
&&{\rm Tr}\{F_1^B(t,0)\rho_0\}\nonumber\\&=&-(i\hbar)^{-1}\int_0^t {\rm Tr}\{[H_1(\tau'),F^H_0(t-\tau')]\rho_0\}d\tau'\nonumber\\
&=&(i\hbar)^{-1}\int_0^t{\rm Tr}\{[H_1(\tau'),\rho_0]F^H_0(t-\tau')\}d\tau'\nonumber\\
&=& {\rm Tr}\{(i\hbar)^{-1}\int_0^t U_0(t-\tau')[H_1(\tau'),\rho_0]U^\dag_0(t-\tau')d\tau' \hspace{0.05cm}F\},
\end{eqnarray}
and
\begin{eqnarray}
\label{2ndexpctation}
&&{\rm
Tr}\{F_2^B(t,0)\rho_0\}\nonumber\\&=&(i\hbar)^{-2}\int_{0}^t
\int_{\tau'}^t {\rm Tr}\{U^{\dag}_0(\tau')
\left[H_1(\tau'),U_0(\tau'-\tau'')[H_1(\tau''), F^H_0(\tau''-t)]U^\dag_0(\tau'-\tau'')\right]U_0(\tau')\rho_0\}d\tau'd\tau'' \nonumber\\
&=&(i\hbar)^{-2}\int_{0}^t \int_{0}^{\tau'}{\rm Tr}\{U^{\dag}_0(\tau'')
\left[H_1(\tau''),U_0(\tau''-\tau')[H_1(\tau'), F^H_0(\tau'-t)]U^\dag_0(\tau''-\tau')\right]U_0(\tau'')\rho_0\} d\tau'd\tau'' \nonumber\\
&=&{\rm Tr}\{(i\hbar)^{-2}\int_0^t\int_0^{\tau'} U_0(t-\tau')\left[H_1(\tau'),U_0(\tau'-\tau'')
[H_1(\tau''),\rho_0]U^\dag(\tau'-\tau'')\right]U^\tau_0(t-\tau') d\tau'd\tau'' \hspace{0.05cm}F\}  ,
\end{eqnarray}
respectively. Two simple formulas ${\rm Tr}\{[A,B]C\}={\rm
Tr}\{[B,C]A\}$ and ${\rm Tr}\{\left[A,[B,C]\right]D\}={\rm
Tr}\{[B,[A,D]C]\} $ are used in the derivations. We immediately
find that the integral terms before the operator $F$ in the
last equalities of these equations are just $\rho_1(t)$ and
$\rho_2(t)$,
respectively~\cite{Kubo,Suzuki}.\\

{\it Conclusion.} we give a simple operator perturbation theory
in the backward Heisenberg picture. Its extension of the classical mechanical systems is straightforward. \\

{\noindent This work was supported in part by the National
Science Foundation of China under Grant No. 11174025. }

\end{document}